\documentclass{article}

\usepackage[cam,center,letter,noinfo]{crop}

\newcommand*\eps{\varepsilon}
\newcommand{\proof}{{\bf Proof:}\ }
\newcommand{\qed}{$\Box$}
\newtheorem{theo}{Theorem}[section]
\newtheorem{lemm}[theo]{Lemma}
\newtheorem{coro}[theo]{Corollary}

\newtheorem{prop}[theo]{Proposition}

\newcommand*\altcropulc{%
\begin{picture}(0,0)
 \unitlength = 1pt
 \thinlines
 \put(-30,0){\circle{10}}
 \put(-30,-5){\line(0,1){10}}
 \put(-35,0){\line(1,0){21}}
 \put(0,30){\circle{10}}
 \put(-5,30){\line(1,0){10}}
 \put(0,35){\line(0,-1){21}}
 \end{picture}
}

\newcommand*\altcropurc{%
 \begin{picture}(0,0)
 \unitlength = 1pt
 \thinlines
 \put(30,0){\circle{10}}
 \put(30,-5){\line(0,1){10}}
 \put(35,0){\line(-1,0){21}}
 \put(0,30){\circle{10}}
 \put(-5,30){\line(1,0){10}}
 \put(0,35){\line(0,-1){21}}
 \end{picture}%
}

\newcommand*\altcropllc{%
 \begin{picture}(0,0)
 \unitlength = 1pt
 \thinlines
 \put(-30,0){\circle{10}}
 \put(-30,-5){\line(0,1){10}}
 \put(-35,0){\line(1,0){21}}
 \put(0,-30){\circle{10}}
 \put(-5,-30){\line(1,0){10}}
 \put(0,-35){\line(0,1){21}}
 \end{picture}%
}

\newcommand*\altcroplrc{%
 \begin{picture}(0,0)
 \unitlength = 1pt
 \thinlines
 \put(30,0){\circle{10}}
 \put(30,-5){\line(0,1){10}}
 \put(35,0){\line(-1,0){21}}
 \put(0,-30){\circle{10}}
 \put(-5,-30){\line(1,0){10}}
 \put(0,-35){\line(0,1){21}}
 \end{picture}%
 }

\cropdef\altcropulc\altcropurc\altcropllc\altcroplrc{altcrop}

\crop[altcrop]
\RequirePackage[dvips]{graphicx}
\RequirePackage{url}
\usepackage{cite}
\RequirePackage{amsfonts}
\RequirePackage{amsmath}
\RequirePackage{amssymb}
\usepackage{float}
\usepackage{listings}
{\makeatletter
 \gdef\xxxmark{%
   \expandafter\ifx\csname @mpargs\endcsname\relax 
     \expandafter\ifx\csname @captype\endcsname\relax 
       \marginpar{xxx}
     \else
       xxx 
     \fi
   \else
     xxx 
   \fi}
 \gdef\xxx{\@ifnextchar[\xxx@lab\xxx@nolab}
 \long\gdef\xxx@lab[#1]#2{{\bf [\xxxmark #2 ---{\sc #1}]}}
 \long\gdef\xxx@nolab#1{{\bf [\xxxmark #1]}}
}

\graphicspath{{images/}}

\begin{document}

\title{Circle Packing for Origami Design Is Hard}
\author{Erik D. Demaine\thanks{
edemaine@mit.edu.
}
\and S\'{a}ndor P. Fekete\thanks{
s.fekete@tu-bs.de.
} \and Robert J. Lang\thanks{
robert@langorigami.com.
}
}
\label{demaine-fekete-lang-chapter}
\date{}

\maketitle

\section{Introduction}
\label{sec:intro}

Over the last 20 years, the world of origami has been changed by the
introduction of design algorithms that bear a close relationship to, if not
outright ancestry from, computational geometry. One of the first robust
algorithms for origami design was the circle/river method (also called the tree
method) developed independently by Lang \cite{l-mafod-94, l-tmood-94,
l-cafod-96} and Meguro \cite{m-jos-92, m-tkr-94}. This algorithm and its
variants provide a systematic method for folding any structure that
topologically resembles a graph theoretic weighted tree. Other algorithms
followed, notably one by Tachi \cite{t-3odbotm-09} that gives the crease
pattern to fold an arbitrary 3D surface.

Hopes of a general approach for
efficiently solving all origami design problems were dashed early on, when Bern
and Hayes showed in 1996 that the general problem of crease assignment ---
given an arbitrary crease pattern, determine whether each fold is mountain or
valley --- was NP-complete \cite{b-otcofo-96}. In fact, they showed more: given
a complete crease assignment, simply determining the stacking order of the
layers of paper was also NP-complete. Fortunately, while crease
assignment in the general case is hard, the crease patterns generated by the
various design algorithms carry with them significant extra information
associated with each crease, enough extra information that the problem of
crease assignment is typically only polynomial in difficulty. This is certainly
the case for the tree method of design \cite{dl-foacaiub-09}.

Designing a
model using the tree method (or one of its variants) is a two-step process: the
first step involves solving an optimization problem where one solves for
certain key vertices of the crease pattern. The second step constructs creases
following a geometric prescription and assigns their status as mountain,
valley, or unfolded. The process of constructing the creases and assigning them
is definitely polynomial in complexity; but, up to now, the computational
complexity of the optimization was not established.

There were reasons for
believing that the optimization was, in principle, computationally intractable.
The conditions on the vertex coordinates in the optimization can be expressed
as a packing problem, in which the packing objects are circles and ``rivers,ее
(which are curves of constant width) of varying size. It is known that many
packing problems are, in fact, NP-hard, and our intuition suggested
that this might be the case for the tree method optimization problem.

In
this paper, we show that this is, in fact, the case. The general tree method
optimization problem is NP-hard. In the usual way with such problems,
we show that any example of {\sc 3-Partition} can be expressed as a tree method
problem. At the same time, we show that deciding whether a given set of circles can
be packed into a rectangle, an equilateral triangle, or a unit square are NP-hard problems,
settling the complexity of these natural packing problems. On the positive side, we show that
any set of circles of total area 1 can be packed into a square of edge length 
$\frac{4}{\sqrt{\pi}}=2.2567\ldots$

\section{Circle-River Design}
\label{sec:design}

\begin{figure}
\centering
\includegraphics[width=4.5in]{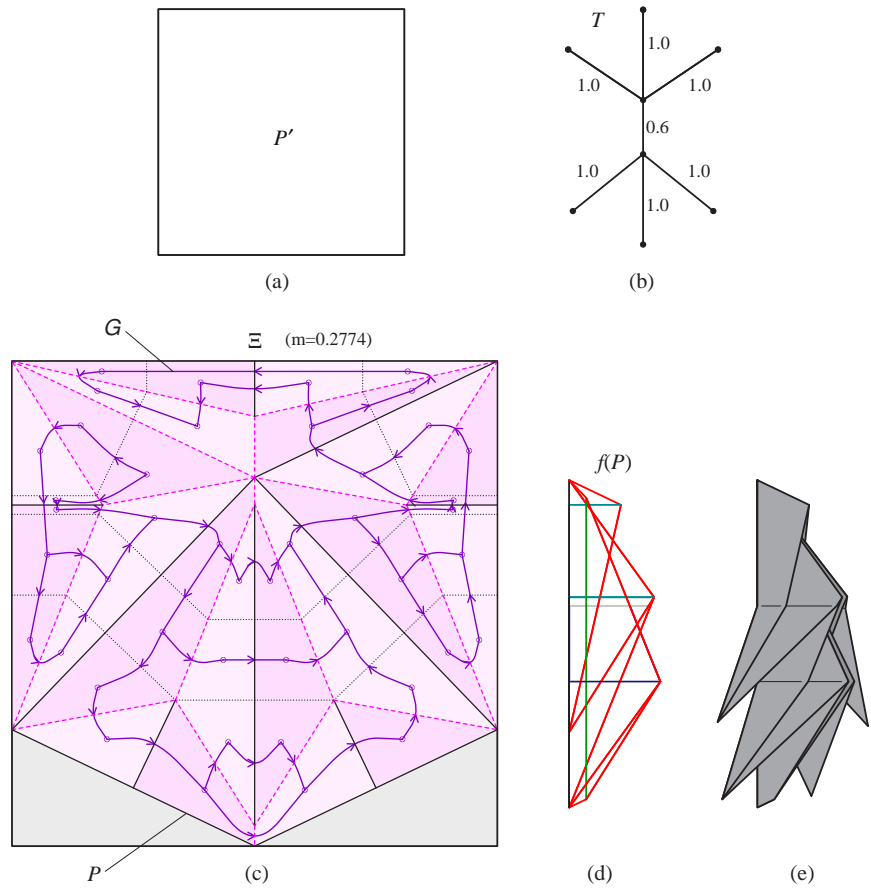}
\caption{
Schematic of the problem. (a) $P'$ is the paper to be folded. (b) $T$ is an edge-weighted tree that describes the desired shape. (c) A solution to the optimization problem, showing creases and the ordering graph on the facets. (d) An x-ray view of the folded form. (e) A visual representation of the folded form.
}
\label{fig:the-problem}
\end{figure}

The basic circle-river method of origami has been described in
\cite{l-cafod-96, dl-foacaiub-09}; we briefly recapitulate it here. As shown in
Figure~\ref{fig:the-problem}, one is presented with a polygon $P'$, which
represents the paper to be folded, and an edge-weighted tree, $T$, which
describes the topology of the desired folded shape. The design problem is to
find the crease pattern that folds $P'$ (or some convex subset) into an origami
figure whose perpendicular projection has the topology of the desired tree $T$
and whose edge lengths are proportional to the edge weights of $T$. The
coefficient of proportionality $m$ between the dimensions of the resulting
folded form and the specified edge weights is called the \emph{scale} of the
crease pattern. The optimization form of the problem is to find the crease
pattern that has the desired topology and that maximizes the scale $m$.

Formally, the problem can be expressed as follows. There is a one-to-one
correspondence between leaf nodes $\{ n_i \}$ of the tree $T$ and
\emph{leaf vertices} $\{ v_i \}$ of the crease pattern whose projections map to
the leaf nodes. We denote the edges of $T$ by $\{ e_j \}$ with edge weights
$w(e_j)$. For any two leaf nodes $n_i, n_j \in T$, there is a unique path
$p_{i,j}$ between them; this allows us to define the \emph{path length}
$l_{i,j}$ between them as
\begin{equation}
l_{i,j} \equiv \sum_{e_k \in p_{i,j}} w(e_k).
\end{equation}

We showed previously \cite{l-cafod-96} that a necessary condition for the existence of a crease pattern with scale $m$ was that for all leaf vertices,
\begin{equation}
\label{eqn:tree-conditions}
| v_i - v_j | \ge m l_{i,j},
\end{equation}
and subsequently, that with a few extra conditions, Equation~\ref{eqn:tree-conditions} was \emph{sufficient} for the existence of a full crease pattern (and we gave an algorithm for its construction). The largest possible crease pattern for a given polygon $P'$, then, can be found by solving the following problem:
\begin{equation}
\label{eqn:optimization-problem}
\mbox{optimize } m \mbox{ subject to } \left\{
\begin{array}{l}
| v_i - v_j | \ge m l_{i,j} \mbox{ for all } i, j \\
v_i \in P' \mbox{ for all } i
\end{array}
\right .
.
\end{equation}

There is a simple physical picture of these conditions: if we surround each vertex by a circle whose radius is the scaled length of the edge incident to its corresponding leaf node and, for each branch edge of the tree, we insert into the crease pattern a curve of constant width (called a river) whose width is the scale length of the corresponding edge, then Equation~\ref{eqn:optimization-problem} corresponds exactly to the problem of packing the circles and rivers in a non-overlapping way so that the centers of the circles are confined to the polygon $P'$ and the incidences between touching circles and rivers match the incidences of their corresponding edges in the tree $T$.

\begin{figure}
\centering
\includegraphics[width=4.5in]{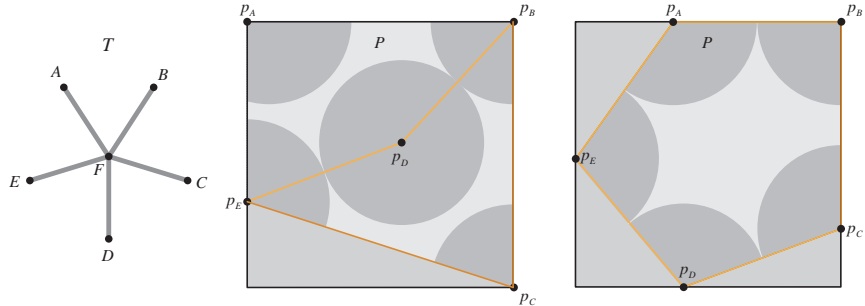}
\caption{
A star tree and two possible solutions for the leaf vertices. Each solution corresponds to a packing of the circles centered on the leaf vertices.
}
\label{fig:tree-feasible-set}
\end{figure}

A special case arises when there are no rivers, i.e., in the case of a star tree with only a single branch node, as illustrated in Figure~\ref{fig:tree-feasible-set}. In this case there are no rivers, and the optimization problem reduces to a single packing of circles, one for each leaf node, whose radius is given by the length of the edge incident to the corresponding node.

Thus, several problems in origami design can be reduced to finding an optimum packing of some number of circles of specified radii within a square (or other convex polygon). Several examples of such problems (and their solutions) are described in \cite{l-ods-03}.

We now show that this circle-packing problem is NP-complete.

\section{Packing and Complexity}
\label{sec:complexity}

Problems of packing a given set of objects into a specific container appear in a large variety of applied and theoretical
contexts. Many one-dimensional variants are known to be NP-complete, e.g., {\sc Bin Packing}, 
where the objective is to pack a set of intervals of given lengths
into as few unit-sized containers as possible. A special case of {\sc Bin Packing} that is still NP-hard
is {\sc 3-Partition}, where an instance is given by $3n$ numbers $x_i$ with $1/4<x_i<1/2$, and $\sum_{i=1}^{3n}x_i=n$.
Clearly, $n$ unit-sized containers suffice for packing the object, iff there is a partition of the $x_i$
into $n$ triples that each have combined weight 1; hence the name {\sc 3-Partition}. An important property of the problem
is that it is {\em strongly} NP-complete: it remains hard even if there is only a constant number of different
values $x_i$ \cite{GareyJ79}.

Like their one-dimensional counterparts, higher-dimensional packing problems tend to be hard. Typically, the difficulty
arises from complicated container shapes (e.g., a non-simple polygon to be filled with a large number of unit 
squares), or complicated objects (e.g., rectangles of many different sizes to be filled into a square, which is a generalization
of {\sc Bin Packing}.)
This does
not mean that packing simple objects into simple containers is necessarily easy: for some such problems it is not even
known whether they belong to the class NP. One example is the problem {\sc Pallet Loading} of deciding whether $n$
rectangles of dimensions $a\times b$ can be packed into a larger rectangle of dimensions $A\times B$, for positive integers
$n, a, b, A, B$: it is open whether the existence of any feasible solution implies the existence of a packing that can
be described in space polynomial in the input size $\log n+\log a+\log b+\log
A+\log B$, as the two different orientations of the small rectangles may give
rise to complicated patterns. (See Problem \#55 in The Open Problems Project,
\cite{cg:DemaineORourke:04}.)

None of these difficulties arises when a limited number of simple shapes without rotation, in particular, different
squares or circles 
are to be packed into a unit square. Leung et al.~\cite{squaresquare}
managed to prove that the problem {\sc
Square Packing} of deciding whether a given set of squares can be packed into a
unit square is an NP-complete problem. Their proof is based on a
reduction of the problem {\sc 3-Partition} mentioned above: any {\sc
3-Partition} instance $\Pi_{3p}$ can be encoded as an instance $\Pi_{sp}$ of {\sc Square Packing},
such that $\Pi_{sp}$ is solvable iff $\Pi_{3p}$ is, and the encoding size of $\Pi_{sp}$ is polynomial in the encoding
size of $\Pi_{3p}$. Membership in NP is not an issue, as coordinates of a feasible packing
are integers of description size polynomial in the encoding size of $\Pi_{sp}$.

In the context of circle/river origami design, we are particularly interested in the problem of {\sc Circle Packing}:
given a set of $n$ circles of a limited number of different sizes, decide whether they can be packed into
a unit square. More precisely, we are interested in {\sc Circle Placement}:
given a set of $n$ circles, place the circle centers on the paper,
such that the overall circle layout is non-overlapping. 
Clearly, this feels closely related to {\sc Square Packing}, so it is natural to suspect
NP-completeness. However, 
  when packing circles, another issue arises: tight packings may give rise to complicated coordinates. 
In fact, the minimum size $C_n$ of a $C_n\times C_n$ square that can accommodate $n$ unit circles is only known
for relatively moderate values of $n$; consequently, the membership of {\sc Circle Packing} in NP is wide open. 
(At this point, $n=36$ is the largest $n$ for which the exact value of $C_n$ is
known; see \cite{specht} for the current status of upper and lower bounds for
$n\leq 10,000$.)

Paradoxically, this additional difficulty has also constituted a major roadblock for establishing
NP-hardness of {\sc Circle Packing}, which requires encoding desired combinatorial structures as 
appropriate packings: this is hard to do when little is known about the structure of optimal packings.

The main result of this paper is to describe an NP-hardness proof of {\sc Circle Placement},
based on a reduction of {\sc 3-Partition}; it is straightforward to see that this also
implies NP-hardness of {\sc Circle Packing}.
 In the following section, we will describe the key idea
of using {\em symmetric 3-pockets} for this reduction: a triple of small {\em ``shim''} circles
$C_{i_1}$, 
$C_{i_2}$, 
$C_{i_3}$ and a medium-sized {\em ``plug''} circles can be packed into such a pocket, 
iff the corresponding triple of numbers 
$x_{i_1}$,
$x_{i_2}$,
$x_{i_3}$ add up to at most 1. In the following sections, we show how symmetric 3-pockets
can be forced for triangular paper (Section~\ref{sec:triangular}), for
rectangular paper (Section~\ref{sec:rectangular}), and for square paper
(Section~\ref{sec:square}).  The technical details for the proof of 
NP-hardness are wrapped up in Section~\ref{sec:gaps} and
Section~\ref{sec:approx}, in which we sketch additional aspects of
filling undesired holes in the resulting packings, 
approximating the involved irrational coordinates, 
and the polynomial size of the overall construction.
On the positive side, we show in Section~\ref{sec:positive} that circle packing becomes a lot easier if one is
willing to compromise on the size of the piece of paper:  we prove that any
given set of circles of total area at most $1$ can easily and recursively be
packed into a square of edge length $\frac{4}{\sqrt{pi}}=2.2567\ldots$

\section{Symmetric 3-Pockets}
\label{sec:pockets}

\begin{figure}
\centering
\includegraphics[width=2.3in]{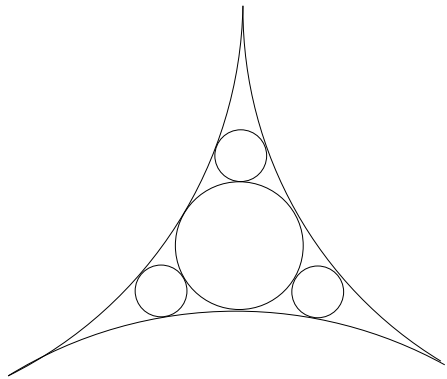}
\caption{
A symmetric 3-pocket with plug and shims.
}
\label{fig:basic0}
\end{figure}

Our reduction is based on the simple construction shown in Figure~\ref{fig:basic0}.
It consists of a {\em symmetric 3-pocket} as the container, which is the area bounded by three
congruent touching circles.

Into each pocket, we pack a medium-sized circle (called a {\em plug})
that fits into the center, and three small identical circles (called {\em shims}) that fit into the
three corners left by the plug. Straightforward trigonometry (or use of Proposition~\ref{prop:formula})
shows that for a pocket formed by three unit circles, 
the corresponding size is $2/\sqrt{3}-1=0.1547...$ for the plug; the value for the shims works out to 
$1/(5+\sqrt{3}+2\sqrt{7+4\sqrt{3}}=0.07044...$.

Clearly, this packing is unique, and the basic layout of the solution does not change when the plug is
reduced in size by a tiny amount, say, $\eps=1/N$ for a suitably big $N$, while each shim is increased by
a corresponding amount that keeps the overall packing tight. This results in a radius of $r_p$ for each
plug, and a radius of $r_s$ for each shim.

Now consider the numbers $x_i$ for $i=1,\ldots,3n$, constituting an instance of {\sc 3-Partition}.
We get a feasible partition iff all triples $(i_1,i_2,i_3)$ are feasible, i.e.,
$\sum_{j=1}^3 x_{i_j} = 1$. By introducing
$x'_i=1/3-x_i$ and using $\sum_{i=1}^{3n} x_i = n$, it is easy to see that a partition is feasible
iff $\sum_{j=1}^3 x'_{i_j}\leq 0$ for all triples $(i_1,i_2,i_3)$. Note that a {\sc 3-Partition} instance
involves only a constant number of different sizes, so there is some $\delta>0$ such that
any infeasible triple $(i_1,i_2,i_3)$ incurs $\sum_{j=1}^3 x'_{i_j}\geq \delta$. By picking $N$ large
enough, we may assume $\delta>\eps$.

\begin{figure}
\centering
\input{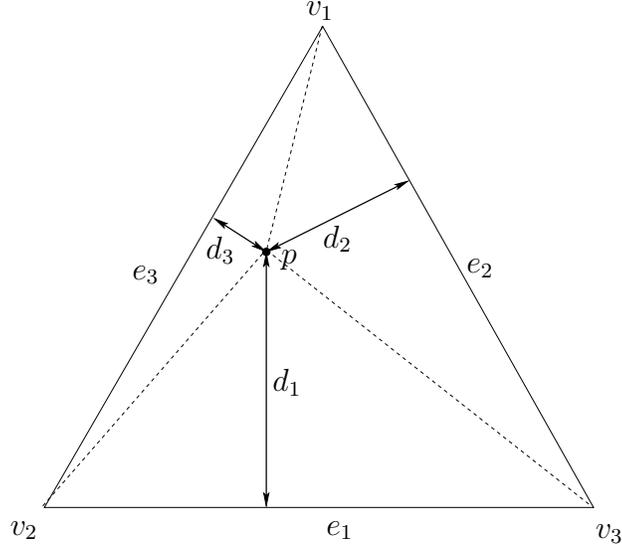}
\caption{
For an equilateral triangle, the sum of distances from the three 
sides is a constant.
}
\label{fig:lemma}
\end{figure}

As a next step, map each $x_i$ to a slightly modified shim $S_i$ by picking the shim radius to be $r_i=r_s-x'_i/N^2$. 
We will make use of the following elementary lemma; see Figure~\ref{fig:lemma}

\begin{lemm}
\label{equilateral}
Consider an equilateral triangle $\Delta=(v_1,v_2,v_3)$ bounded  by the lines $\ell_1, \ell_2, \ell_3$ through
the triangle edges $e_1, e_2, e_3$. For an arbitrary point $p$, let
$d_j$ be the distance of $p$ from $\ell_j$. Define $y_j=d_j$, if $p$ is on the same side
of $\ell_j$ as $\Delta$, and $y_j=-d_j$ if $p$ is separated from $\Delta$ by $\ell_j$.
Then $\sum_{j=1}^3 y_j$ is independent of the position of $p$.
\end{lemm}

\proof
Consider the three triangles 
$(v_1,v_2,p)$,
$(v_2,v_3,p)$,
$(v_3,v_1,p)$. Their areas are 
$d_3/2$,
$d_1/2$,
$d_2/2$, hence $y_1/2+y_2/2+y_3/2$ is always equal to the area of $\Delta$,
i.e., a constant.
\qed

The crucial argument for our reduction is the following.

\begin{lemm}
\label{le:reduce}
A set of three shims 
$S_{i_1}$,
$S_{i_2}$,
$S_{i_3}$ and a plug $P$ of radius $r_p$ can be packed into a 3-pocket, 
iff $\sum_{j=1}^3 x'_{i_j}\leq 0$, i.e., if $(i_1,i_2,i_3)$ is feasible.
\end{lemm}

\begin{figure}
\centering
\input{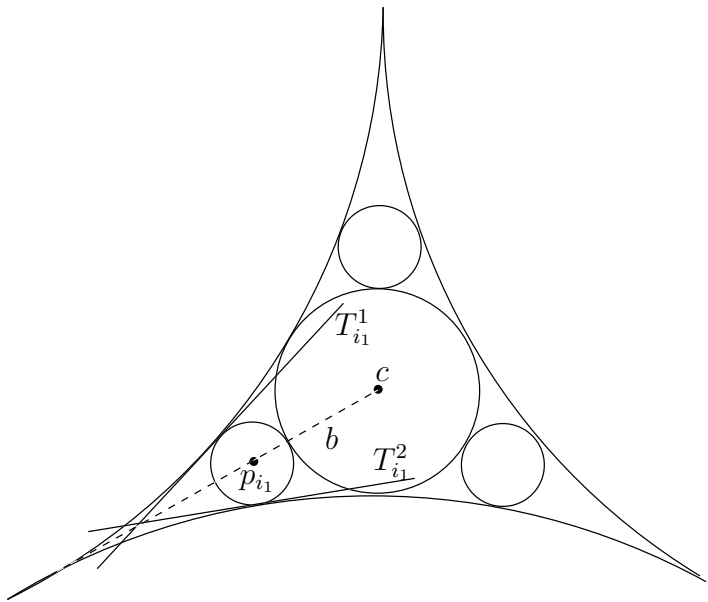}
\caption{
Changing the size of a shim.
}
\label{fig:basic}
\end{figure}

\proof
Refer to Figure~\ref{fig:basic}. Let $c$ be the center point of the pocket.
For each of the three corners of the pocket, consider
the two tangents $T^1_{i_j}$  and $T^2_{i_j}$
between an unmodified shim of radius $r_s$ and the touching pocket boundary; let $2\phi\in]0,\pi[$ be the
angle enclosed by those two tangents. (If the pocket was an equilateral triangle,
we would get $\phi=\pi/6$; the exact value for pockets with circular boundaries can be computed,
but the exact value does not matter.)

Now consider the shim motion arising by modifying $r_s$ by $x'_{i_j}/N^2$, while keeping the shim tightly
wedged into the corner. This moves its center point $p_{i_j}$ along the bisector $b$ between
$T^1_{i_j}$  and $T^2_{i_j}$. Let $c=1/\sin\phi$. Considering the first-order expansion
of the shim motion, we conclude that $p_{i_j}$ moves by $c\times x'_{i_j}/N^2+\Theta(1/N^4)$ along
$b$, to a position $q_j$. 

\begin{figure}
\centering
\input{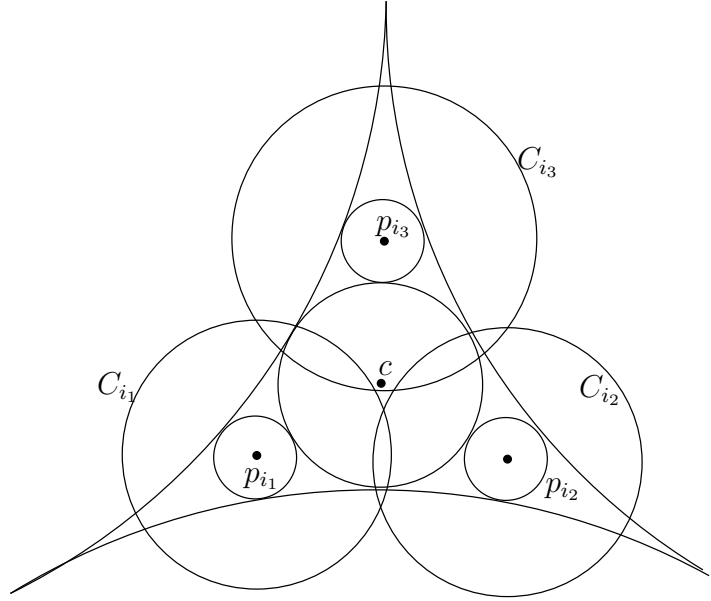}
\caption{
Finding a feasible placement for the plug.
}
\label{fig:basic2}
\end{figure}

\begin{figure}
\centering
\input{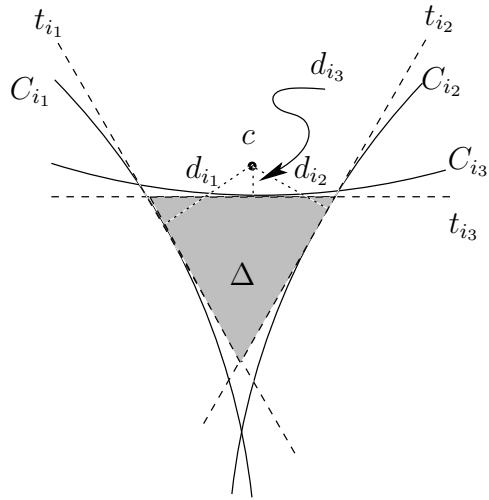}
\caption{
The existence of a feasible placement for the plug depends on the sum
of distances of $c$ from the sides of $\Delta$. (Distances are not drawn to scale so that circles and tangents can be distinguished; in reality, they are 
much closer.)
}
\label{fig:basic3}
\end{figure}

Finally, refer to Figure~\ref{fig:basic2} and consider the possible placement of a plug after placing the modified shims $S_{i_1}$,
$S_{i_2}$, $S_{i_3}$ into the the three corners; this requires finding a point within the pocket
that has distance at least $r_p+r_s-x'_{i_j}/N^2$ from each $p_{i_j}$.
For this purpose, consider the circle $C_{i_j}$ of radius $r_p+r_s-x'_{i_j}/N^2$ around each
$p_{i_j}$. As shown in Figure~\ref{fig:basic3},
let $t_{i_j}$ be the tangent to $C_{i_j}$ at the point closest to $c$; let 
$d_{i_j}$ be the distance of $c$ to $t_{i_j}$. If we define 
$y_{i_j}=d_{i_j}$ for $c$ is outside of $C_{i_j}$, and
$y_{i_j}=-d_{i_j}$ for $c$ is inside of $C_{i_j}$, then 
$y_{i_j}= -((c+1)\times x'_{i_j}/N^2+\Theta(1/N^4))$.
Now consider the set $\Delta$ of points separated by 
$t_{i_1}$ from $p_{i_1}$,
$t_{i_2}$ from $p_{i_2}$,
$t_{i_3}$ from $p_{i_3}$.
Making use of Lemma~\ref{equilateral}, we conclude that $\Delta$ is a nonempty isosceles
triangle, iff $\sum y_{i_j} \geq 0$, i.e., iff $\sum x'_{i_j}/N^2 -\Theta(1/N^4) \leq 0$. 
Given that $\sum x'_{i_j}>0$ implies $\sum x'_{i_j}\geq \delta > 1/N$, we conclude that 
$\sum x'_{i_j}\leq 0$ implies the existence of a feasible packing. 

Conversely, consider $\sum x'_{i_j}>0$. Given that each $t_{i_1}$ has a distance $\Theta(1/N^2)$
from $c$, we observe that the corners of the triangle formed by $t_{i_1}$, 
$t_{i_2}$, $t_{i_3}$ are within $\Theta(1/N^4)$ from 
$C_{i_1}$, $C_{i_2}$, $C_{i_3}$. However, because of $\sum x'_{i_j}\geq \delta > 1/N$,
we conclude that any point of $\Delta$ is at least $\Theta(1/N^3)$ from being feasible.
This implies that there is no feasible placement for the plug, concluding the proof.

\qed

\section{Triangular Paper}
\label{sec:triangular}

For making use of Lemma~\ref{le:reduce} and completing the reduction, we need to define a set of 
circles (called {\em rocks}) that can only be packed in a way that results in a suitable number of
3-pockets. In the case of triangular paper, this is relatively easy by making use of
a result by Graham~\cite{graham}.

\begin{prop}
An equilateral triangle of edge length $2k$ has a unique
packing of $(k+2)(k+1)/2$ unit circles; this uses a hexagonal grid pattern,
placing circles on the corners of the triangle.
\end{prop}

\begin{figure}
\centering
\includegraphics[width=2.5in]{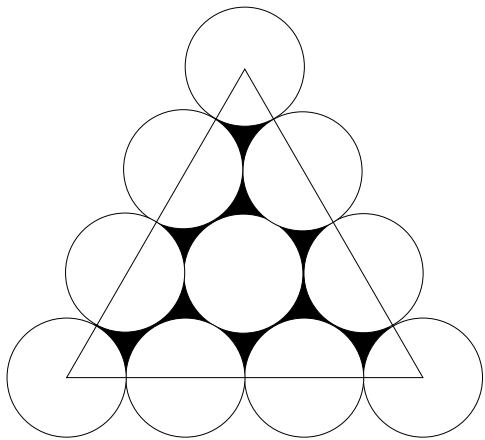}
\caption{
The unique packing of $(k+2)(k+1)/2$ unit circles into 
an equilateral triangle of edge length $2k$ leaves $k^2$
identical symmetric 3-pockets.
}
\label{fig:triangle}
\end{figure}

This creates $\sum_{i=1}^{k}(2i-1)=k^2$ symmetric 3-pockets. After handling some issues of
accuracy and approximation (which are discussed in Section~\ref{sec:approx}), we get the desired
result.

\begin{theo}
Circle/river origami design for triangular paper is NP-hard.
\end{theo}

As a corollary, we get

\begin{coro}
It is NP-hard to decide whether a given set of circles can be packed into
an equilateral triangle.
\end{coro}

\section{Rectangular Paper}
\label{sec:rectangular}

\begin{figure}
\centering
\includegraphics[width=2.5in]{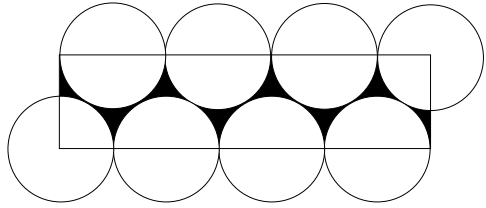}
\caption{
Packing $2k$ unit circles into 
a rectangle of dimensions $2k-1$ and $\sqrt{3}$ leaves $2k-2$
identical symmetric 3-pockets.
}
\label{fig:rectangle}
\end{figure}

Similar to triangular paper, it is easy to force a suitable number of symmetric
3-pockets for the case of rectangular paper, see Figure~\ref{fig:rectangle}:
disregarding symmetries, $2k$ unit circles can only be packed into an
$2k-1$ by $\sqrt{3}$ rectangle in the manner shown. This creates
$2k-2$ symmetric 3-pockets, which can be used for the hardness proof.

Because the input created for encoding an instance $\Pi_{3p}$
of {\sc 3-Partition} needs to be a set of rationals whose size is
bounded by a polynomial in the encoding size of $\Pi_{3p}$, the irrational
numbers needs to be suitably approximated without compromising the
overall structure. This will be discussed in Section~\ref{sec:approx}. As a consequence,
we get

\begin{theo}
Circle/river origami design for rectangular paper is NP-hard.
\end{theo}

This yields the following easy corollary.

\begin{coro}
It is NP-hard to decide whether a given set of circles can be packed into
a given rectangle.
\end{coro}

\section{Square Paper}
\label{sec:square}

Setting up a sufficient number of symmetric 3-pockets for square paper is slightly trickier:
there is no infinite family of positive integers $n$, for which 
the optimal patterns of packing $n$ unit circles into a minimum-size square
are known. As a consequence, we make use of a different construction;
without loss of generality, our piece of paper is a unit square.

\begin{figure}
\centering
\includegraphics[width=2.5in]{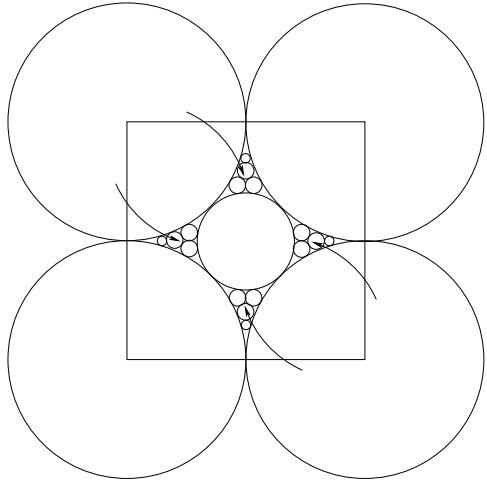}
\caption{
A gadget for creating identical triangular pockets:
the shown set of 13 circles has a unique packing into a symmetric
4-pocket. This creates four smaller symmetric 3-pockets, indicated by arrows.
}
\label{fig:4-pocket}
\end{figure}

As a first step, we use four large circles of radius $1/2$, creating
a symmetric {\em 4-pocket}, as shown in Figure~\ref{fig:4-pocket}.
Now a circle of radius $\frac{\sqrt{2}-1}{2}$ has a unique feasible placement
in the center of the pocket, leaving four smaller auxiliary pockets, as shown.

Now we use 12 identical ``plug'' circles and four slightly smaller ``fixation''
circles, such that three plugs and one shim have a tight packing as shown
in the figure. For these it is not hard to argue that not more than three
plugs fit into an auxiliary pocket, ensuring that precisely three must be placed
into each pocket. Moreover, it can be shown that at most one additional
shim can be packed along with the three plugs; this admits precisely the
packing shown in the figure, creating a symmetric 3-pocket in each auxiliary pocket.
In addition, we get a number of undesired asymmetric pockets, which must be used
for accommodating appropriate sets of ``filling'' circles, leaving only small gaps that cannot
be used for packing the circles that are relevant for the reduction.

\begin{figure}
\centering
\includegraphics[width=2.5in]{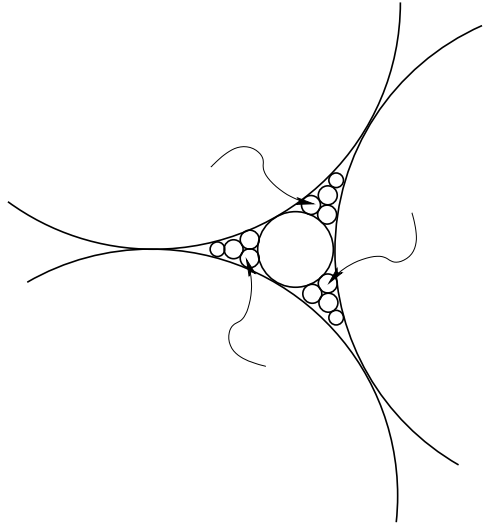}
\caption{
A gadget for creating multiple identical triangular pockets:
the shown set of 13 circles has a unique packing into a symmetric
3-pocket. This creates three smaller symmetric 3-pockets, indicated by arrows.
}
\label{fig:3-pocket}
\end{figure}

As shown in Figure~\ref{fig:3-pocket}, we can use a similar auxiliary
construction (consisting of 13 circles) for the 3-pockets in a recursive
manner in order to replace each symmetric 3-pocket by three smaller symmetric 3-pockets.
The argument is analogous to the one for 4-pockets. Again, additional filling circles are used;
these do not compromise the overall structure of the packing, as the overall
argument holds 

\begin{theo}
Circle/river origami design for square paper is NP-hard.
\end{theo}

This yields the following easy corollary.

\begin{coro}
It is NP-hard to decide whether a given set of circles can be packed into
a given square.
\end{coro}

\section{Filling Gaps}
\label{sec:gaps}

The above constructions create a number of additional gaps in the form of asymmetric 3-pockets.
Each is bounded by three touching circles, say, of radius $r_1$, $r_2$, $r_3$. By adding appropriate sets of
``filler'' circles that precisely fit into these pockets, we can ensure that they cannot be exploited
for sidestepping the desired packing structure of the reduction. Computing the necessary radii can simply be done
by using the following formula.

\begin{prop}
\label{prop:formula}
The radius $r$ of a largest circle inscribed into a pocket formed
by three mutually touching circles with radii $r_1$, $r_2$, $r_3$ satisfies
\[1/r= 1/r_1+1/r_2+1/r_3+2\sqrt{1/r_1r_2+1/r_1r_3+1/r_2r_3}\]
\end{prop}

Note that the resulting $r$ is smaller than the smallest $r_i$, and at least a factor of 3 smaller than
the largest of the circles. Therefore, computing the filler circles
by decreasing magnitude guarantees that all gaps are filled in the desired fashion, and that
only a polynomial number of such circles is needed.

\section{Encoding the Input}
\label{sec:approx}

In order to complete our NP-hardness proof for {\sc Circle Placement}, we still need to 
ensure that the description size of the resulting {\sc Circle Placement} instance is polynomial
in the size of the input for the original {\sc 3-Partition} instance. It is easy to see from
the above that the total number of circles remains polynomial. This leaves the issue of
encoding the radii themselves: if we insist on tightness of all packings, we get irrational numbers
that can be described as nested square roots. As described in Section~\ref{sec:pockets},
the key mechanisms of our construction still work if we use a sufficiently close
approximation. This allows to use sufficiently tight approximations of the involved
square roots in other parts of the construction, provided the involved computations are fast and easy
to carry out. For our purposes, even Heron's quadratically converging method (which doubles the number of correct 
digits in each simple iteration step) suffices.

\section{A Positive Result}
\label{sec:positive}

Our NP-hardness results imply that there is little
hope for a polynomial-time algorithms that computes the smallest possible
triangle, rectangle or square for placing or packing a given set of
circles. However, it is possible to guarantee the existence of a feasible
solution, if one is willing to use larger paper. In fact, we show that
a square of edge length $4/\sqrt{\pi}=2.2567...$ suffices for packing any
set of circles that have total area 1.

\begin{theo}
Consider a set ${\cal S}$ of circles of total area 1, and a square $S$ of 
edge length $4/\sqrt{\pi}$. Then  ${\cal S}$ can be packed into $S$.
\end{theo}

\begin{figure}
\centering
\includegraphics[width=2.5in]{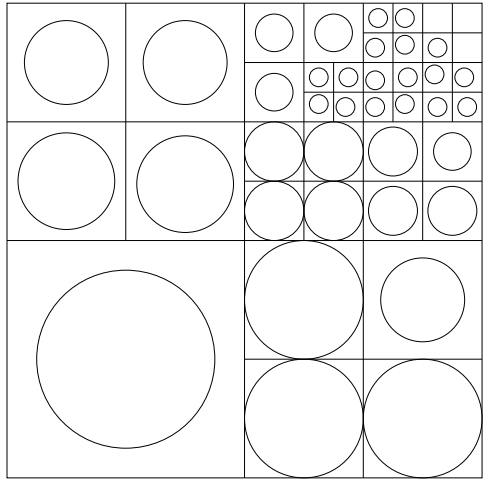}
\caption{
A quad-tree packing guarantees that any set of circles of total area at most 1 can be packed into
a square of edge length $\gamma=4/\sqrt{\pi}=2.2567...$
}
\label{fig:quadtree}
\end{figure}

\proof
Refer to Figure~\ref{fig:quadtree}. For each circle $C_i$ of radius $r_i$,
let $n_i$ be chosen such that $\frac{\gamma}{2^{n_i+1}}<r_i\leq \frac{\gamma}{2^{n_i}}$. Hence, replacing each $C_i$ by a square $S_i$ of size
$\frac{\gamma}{2^{n_i}}$ increases the edge length by a factor of at most
$\gamma=4/\sqrt{\pi}$. Now a recursive subdivision of $S$ into sub-squares
of progressively smaller size can be used to pack all 
squares $S_i$, showing that all circles $C_i$ can be packed.

\section{Conclusions}
\label{sec:conc}

In this paper, we have proven that even the aspect of circle packing
in circle/river origami design is NP-hard. On the positive side,
we showed that the size of a smallest sufficient square for
accommodating a given set of circles can easily be approximated within
a factor 2.2567... 
A number of interesting open questions remain:

\begin{figure}
\centering
\includegraphics[width=2.5in]{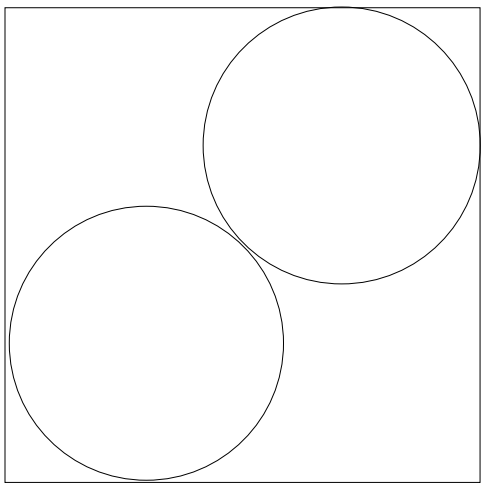}
\caption{
A lower-bound example for packing circles: two circles of area 1/2 require
a square of edge length at least 1.362...
}
\label{fig:worst}
\end{figure}

\begin{itemize}
\item Our 2.2567-approximation is quite simple. The performance guarantee
is based on a simple area argument. This gives rise to the following
question: what is the smallest square that
suffices for packing any set of circles of total area 1? We believe the
worst-case may very well be shown in Figure~\ref{fig:worst}, which
yields a lower bound of $(1+\sqrt{2})/\sqrt{\pi}=1.362\ldots$ There
are ways to improve the upper bound; at this point, we can
establish $2\sqrt{2}/\sqrt{\pi}=1.5957\ldots$ \cite{df-wbdp-10}.
\item The same question can be posed for {\em placing} circles instead
of packing them.
\item The approximation of circle packing does not produce a ``clustered''
layout as required by circle/river origami design, where 
objects that are close in the hierarchy should be place in close vicinity.
In the absence of rivers,
we can reproduce the quad-tree packing in this context by making use
of a space-filling curve. 
\item On the other hand, we do not know yet how
to approximate the necessary paper size in the presence of rivers of
positive width.
\end{itemize}

\section*{Acknowledgments}

We thank Ron Graham for several helpful hints concerning the state of the
art on packing circles. We also thank Vinayak Pathak for pointing out a
numerical typo related to Figure~\ref{fig:worst}.

\bibliographystyle{abbrv}
\bibliography{rjl-origami,safe}

\end{document}